# Amino acid frequency and domain features serve well for random forest based classification of thermophilic and mesophilic protein; a case study on serine proteases

Jithin S. Sunny[a], Lilly M. Saleena*[a]

[a]Department of Biotechnology, SRM Institute of Science and Technology, SRM Nagar, Kattankulathur, 603203, Kanchipuram, Chennai TN, India

*Corresponding author: saleenam@srmist.edu.in

**Abstract**

Thermostability is an important prerequisite for enzymes employed for industrial applications. Several machine learning based models have thus been formulated for protein classification based on this particular trait. These models have employed features derived from sequences, structures or both resulting in a >93% accuracy based on a 10-fold cross-validation. Besides using various proteins from a wide range of organisms, such studies also rely on hundreds of features. In the present study, an enzyme specific classification model was created using significantly less number of features that provides a similar accuracy of classification for thermophilic and non-thermophilic enzyme serine proteases. For building the classifier, 219 thermophilic and 200 mesophilic bacterial genomes were mined for their respective serine protease sequences. Features were extracted for 800 sequences followed by feature selection. We deployed a random forest based classifier that identified thermophilic and non-thermophilic serine proteases with an accuracy of 95.71%. Knowledge of thermostability along with amino acid positional shifts can be vital for downstream protein engineering techniques. Thus, to emphasize the real time application of the enzyme specific classification model, a web platform has been designed. Combining the sequence data and the classification model, this prototype can allow users to align their query serine protease sequence against the custom database and identify its thermophilic nature.

**Keywords**: Serine protease, Thermophilic, Random Forest, Web tool

# 1. Introduction

Data-driven approaches for screening specific enzyme properties are being constantly developed. These methods require information derived from sequences such as co-evolution, substitutions, etc., to develop an accurate predictive model [1]. This method of quantifying the behavior of the data in-hand can be undertaken by machine learning (ML). A ML approach uses knowledge from the information-rich protein sequences and guide targeted protein engineering endeavors. Naturally found enzymes or wild type enzymes as they are known, function optimally in their actual environment. This limits their use in specific applications desired for industrial or scientific purposes. A combined approach involving ML and experimental techniques can be applied to address this limitation. Two prominent experimental strategies are rational design and directed evolution [2]. While directed evolution requires high throughput screening which in-turn demands large resources and intensive time scales, rational design is limited by the large gap that currently exists between sequence and structure knowledge availability. By combining ML based predictions, valuable prerequisites can be included in the above two enzymes engineering approaches.

The predictions that ML based approaches provide are based on several enzyme properties amongst which thermostability is of prime importance [3]. An enzyme is primarily a sequence of amino acids that are arranged in a specific order. Since this order also determines their thermophilic nature, knowledge of sequences can be integrated to build accurate models which can then fetch useful predictions for enzyme engineering processes [4, 5]. Over the past, protein classification based on this property has been performed using multiple machine learning methods such as support vector machines, deep neural networks and especially random forest (RF) classifiers [6]. While Wu et al [7] used 111 protein features based on primary secondary and tertiary structural parameters, Zuo et al [8] used only amino acid content as an input feature for a dataset of more than 3000 thermophilic protein from various sources. Similarly, Hao et al [9] used 900+ sequences and used the popular amino acid composition and dipeptide features. All the studies reported >90% accuracy for their classification models. These classifiers have been trained on hundreds of diverse thermophilic proteins from various species belonging to Archaea, Bacteria and other vertebrates and invertebrates. Feature extraction in such processes is primarily based on amino acid composition and physicochemical properties imparted by the amino acid such as hydrophobicity, polarity, etc. Several studies conducted across multiple species have revealed unique amino acid level signatures. Residues such as Val, Ile, Glu, Arg were observed to be higher, and Gly, Met, Gln were lower [10]. Besides these residues, a higher occurrence of salt-bridges is also observed [11, 12]. Other important parameters include protein domains, protein binding regions, evolutionary residue conservation of residues, relative position of amino acids amongst others. Domain annotations of proteins have revealed conserved regions of proteins. These regions have often been shown to possess more theoretical sensitivity when it comes to identification than the sequence similarity strategy [13]. Characteristics from the protein binding region can be another significant feature. A protein complex may be composed of multiple binding regions which is dependent on the residue atoms of the interface [14] which is yet another determining factor for the protein functioning. Similarly, a wide range of conservation scores can be given to each residue in a protein which can be an important measure for its function [15]. Such parameters may have a significant role in determining the predictions of thermostability.

In the present study we aim to introduce an enzyme specific thermophilic classification model using the above discussed features. We use amino acid composition along with iso-electric point and several domain features to build a RF classifier. RF has been previously tested in similar works yielding highly accurate classification models. We lay out a workflow for a bacterial serine protease classification model (**Fig 1**). Serine proteases (SPs) are alkaline proteases, a hydrolase driven by nucleophilic core with a wide range of applications in industry [16]. SPs are mostly isolated from bacterial species and based on their source, they may display a variety of features. In the present study we make use of this diversity to analyze the enzyme further. Serine proteases isolated from thermophilic bacterial species display high resistance to temperatures above 60°C making it suitable for a variety in industrial processes [17, 18]. Establishing a classification model for this protein can serve both rational and directed evolution protein engineering endeavors. Moreover, this workflow can be utilized for other enzyme classes involving thermostability as classifying criteria.

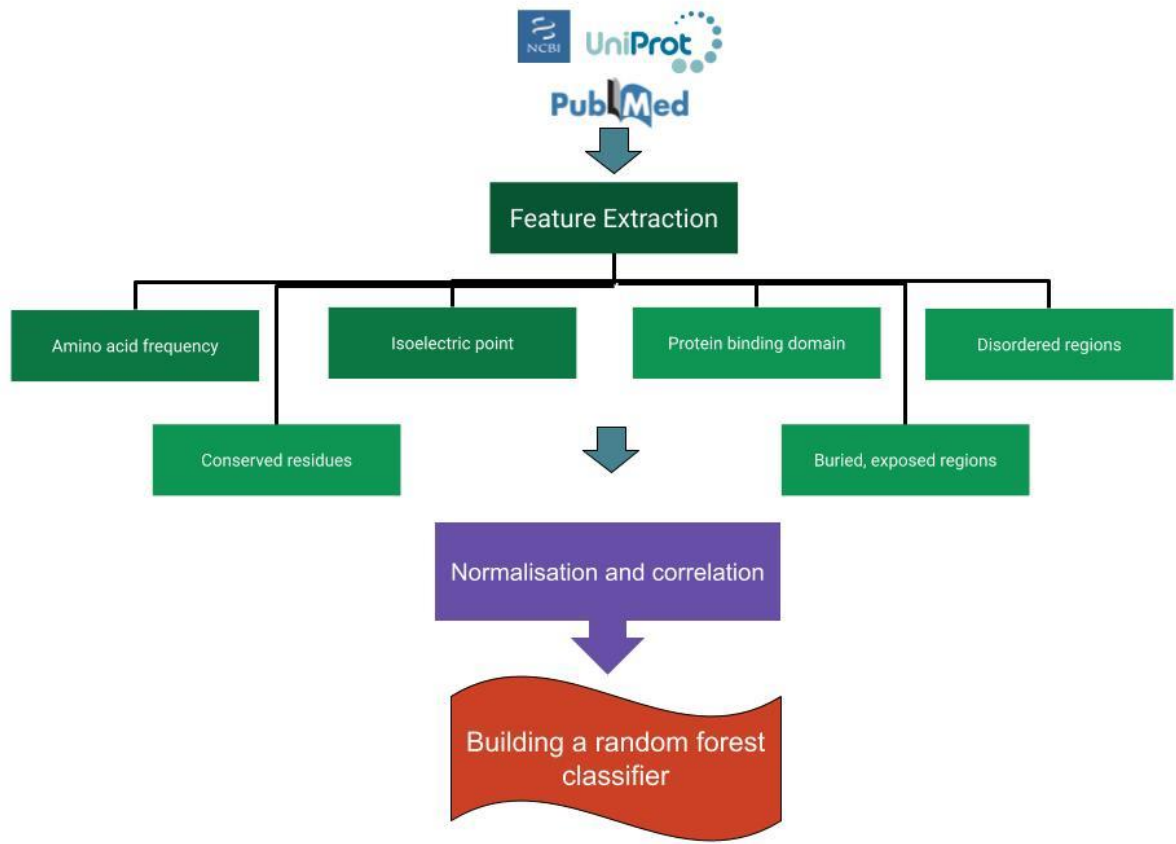

**Fig 1. Workflow for enzyme specific classifier**

## 2. Materials and Methods

### 2.1 Protein sequence mining

Initially, online repositories such as NCBI [19], PDB [20] and other biological databases were searched for SPs sequences from thermophilic bacterial genomes. First, thermophilic bacterial genomes were downloaded from NCBI which were later uploaded into RAST annotation server [21]. For manual curation of SPs, an Entrez application programming interface (API) was used [22]. An in-house python script was used to assist manual curation by integrating PubMed search with ESearch API. Additionally, relevant literature with the enzymes was searched manually also (**Fig 2**). The sequence data was curated in fasta format. These stored sequences were processed using BLAST+ standalone tool to generate a custom database. Non-thermophilic SPs were also extracted from NCBI protein database.

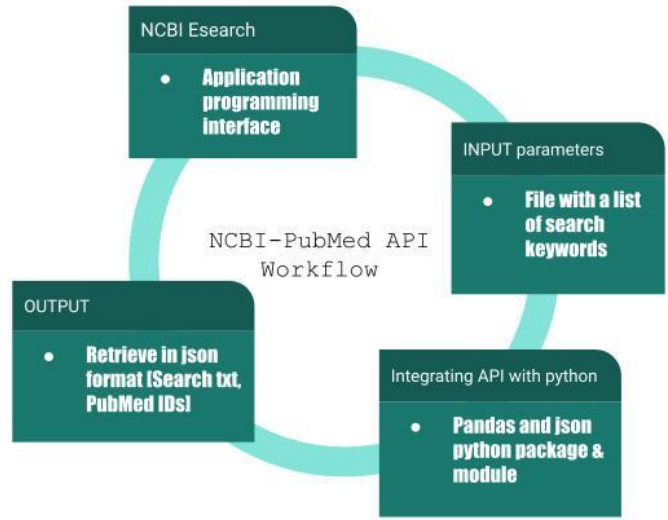

**Fig 2. Literature review pipeline**

## 2.2 Feature extraction

Amino acid percentage was extracted from both thermophilic and non-thermophilic SPs. These frequencies of 20 amino acids were stored for all the SPs. Next, the pI for each sequence was evaluated using the pI calculator from BioPerl [23]. The third set of features we used was the number of residues belonging to the protein domain region. This prediction was based on InterPro results. Domain ranges were fetched for both thermophilic and non-thermophilic SPs using this online tool. Next, the number of protein binding sites for each sequence was predicted using the online tool PredictProtein [24]. Besides protein binding, this tool also predicts disordered regions, conserved residues, buried and exposed regions. The number of these regions spanning each sequence was used as a feature set. To remove the redundant information from the feature set, Pearson's correlation was performed. First, the normalization was performed using the equation:

$$x_i^n = \frac{x_i - \bar{x}}{x_{max} - x} \quad (1)$$

In the above equation, $x_i$ represents the *i*-th value in the feature set. $x_{max}$ and $x$ represents the maximum and minimum values of the feature set. Following normalization, Pearson's correlation was used for estimating correlations between feature pairs.

## 2.3 Building a classifier

Next, a training set was generated using the SPs sequences retrieved above and their mesophilic counterparts. Labels 1 (thermophilic) and 0 (non-thermophilic) were placed for each sequence. For training and prediction, the extracted features were provided to the random forest model. The python library Sklearn was used for running the classifier. The standard random forest generates a collection of decision trees by randomly selecting subspaces from different sets of training data. The final prediction from the model is determined by majority vote.

## 2.4 Evaluating the model performance

Sensitivity (*Se*), Specificity (*Sp*) and accuracy (*Acc*) were used to evaluate the performance of the classifier.

$$Se = \frac{TP}{TP+FN} \times 100\% \quad (2)$$

$$Sp = \frac{TN}{TN+FP} \times 100\% \quad (3)$$

*Se* and *Sp* represents the performance of the classifier. *Se* represents the number of SPs sequences identified as thermophilic. Here, TP represents the thermophilic SPs and FN is the number of thermophilic SPs predicted as non-thermophilic. Similarly, TN represents a number of non-thermophilic SPs. FP shows the number of non-thermophilic SPs identified as thermophilic. The third parameter was $Acc$.

$$ACC = \frac{TP+TN}{TP+TN+FP+FN} \quad (4)$$

The accuracy reflects the ability of prediction of the RF classifier using the test-set.

## 2.5 Developing a model protein engineering platform

We built a prototype sequence based classifier tool of heat stable enzyme for SPs. The two features of the tool are SPs BLAST and the built-in classifier. The SPs sequences retrieved from the thermophilic sequences were first used to create a custom database which later was assigned to the tool. The user provided sequence can be submitted and the resulting alignment with top 10 thermophilic SPs within the cut-off e-value, percentage identity and other alignment parameters can be visualized in a separate output page. The second tool is the classifier itself. Similar to the BLAST input, users can submit their SP sequence to identify the thermophilic nature.

## 3. Results and Discussion

Thermophilic bacterial genomes from NCBI were annotated and the protein files were analyzed. Manual curation also produced SPs reported from experimental studies. The Esearch API generated PubMed IDs for the corresponding SPs study. The output was parsed and the protein sequences were extracted. A total of 470 SPs sequences were extracted from 291 thermophilic bacterial species and literature survey (Supplementary file 1). Similarly, 300 non-thermophilic SPs sequences were also downloaded from NCBI. All the sequences were retrieved in fasta format.

Pearson's correlation revealed 3 amino acids ALA, GLY and VAL to be positively correlated (0.71). Only GLY was retained in the feature set. None of the physicochemical properties; domain range, protein binding site residues, isoelectric point (pI), conserved residues were observed to be correlated. The reduced feature set tested with a random forest algorithm showed more than 90 % value for all three indicators. The accuracy indicators Se, Sp and ACC were

95.53%, 96.01% and 95.71% respectively. The model under study has comparatively higher accuracy than previous studies that have performed classification of thermophilic and non-thermophilic protein. It can be observed that enzyme specific data could require less features for classification based on thermophilic nature.

Amino acid frequency, pI, number of residues in each domain, residues with lowest evolutionary conservation scores, number of protein binding regions, exposed and buried regions formed the classification features. Parameters for almost all studies till date have utilized amino acid composition, multiple sequence alignment profiles, pseudo amino acid composition [25]. These studies involve prediction and classification problems involving proteins with significantly less homology. With the 24 features the RF presented accuracy comparable to the RF classifier of Changli et al. which is 95.69% [11]. In perhaps one of the most accurate study yet, they tested thermophilic proteins belonging to a wide range of organisms. Similarly, the decision tree used by Li et al. [7] involved 111 features for 580 thermophilic proteins and their study involved both sequence and structural data as a feature set. Another major study in this field was conducted by Yong et al., and Hao et al. While a K-nearest neighbor classifier had an accuracy of 91% Support vector machine yielded 93.27% accuracy (**Fig 3**). All of the above studies base their classification on various classes of proteins belonging to different species. The present study however tries to showcase that a specific classification model built using only serine proteases requires significantly less number of features to achieve similar accuracy.

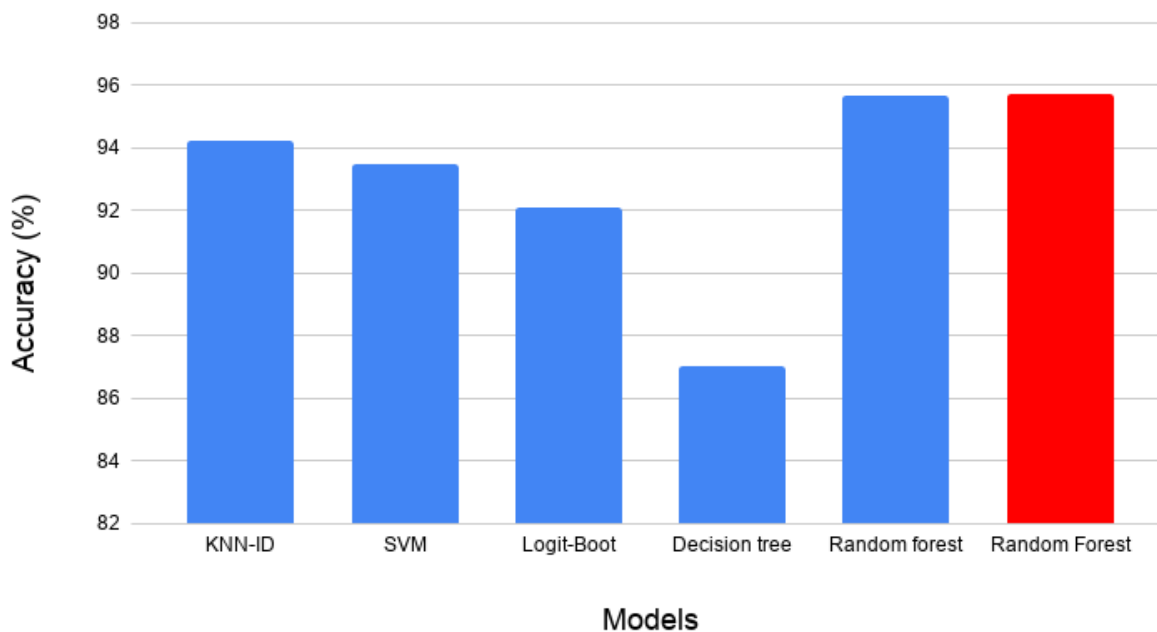

**Fig. 3** Comparative accuracy for Machine learning algorithms used for thermophilic and non-thermophilic protein classification

**Web tool prototype**

First, the BLAST tool allows users to align their SP sequence against the in-built library of 400+ sequences. The easy access allows the user to submit and retrieve the results in a separate page with sequences alignment and their corresponding scores. Since thermophilic nature is highly dependent on the sequence, this alignment feature can be very useful for designing user-defined sequences. The alignment tool can further aid the users in finding similar SPs to their query. Thus unknown/novel proteins showing SP behavior can be identified using this interface. Second and perhaps the significant part of this platform is the classifier interface. The thermophilic property of SP is heavily dependent on the pattern of amino acid arrangement. This exact property is being used by the classifier to predict whether the input SP sequence exhibits a thermophilic nature or not. Upon submitting the SP sequence, the RF classifier will generate the result as a binary classification of either "0" or "1", where "1" shows that the sequence provided is indeed thermophilic in nature. The built-in classifier currently predicts with an accuracy of 95.71%. **Fig 4** Shows our proposed web page design for the tool.

**Fig. 4 Prototype of thermophilic Serine protease analysis tool**

## 4. Conclusion

Thermophilic serine protease is a highly desired enzyme that is required for various industrial applications. Hence, availability of a unified resource platform for its sequence analysis can prove vital. To engineer thermostable enzymes for industrial usage, there exist a variety of pre-requisites. By combining a machine learning approach we have successfully built an accurate model to classify serine protease based on its thermophilic nature. Further, the same study model can be applied to any class of enzyme for its protein engineering studies.

## Acknowledgements

The authors are thankful to SRM-IST dept. of Biotechnology for their support

## Conflict of interest

Authors declare no competing interests.

## Author's contributions

Jithin S. Sunny: Data collection, analysis, manuscript writing

Lilly M. Saleena: Conceptualization, Data collection, Review

# Supplementary File 1

Below is the list of thermophilic bacterial species whose genomes were analyzed

| | | | | | | | |
|---|---|---|---|---|---|---|---|
| *Acetomicrobium faecale* | *Anoxybacter fermentans* | *Bacillus thermotolerans* | *Alicyclobacillus pomorum* | *Coprothermobacter platensis* | *Fervidobacterium thailandense* | *Kyrpidia tusciae* | *Thermus aquaticus* |
| *Acetomicrobium flavidum* | *Apibacter mensalis* | *Bacillus vallismortis* | *Alicyclobacillus sacchari* | *Coprothermobacter proteolyticus* | *Fontimonas thermophila* | *Laceyella sacchari* | *Thermus arciformis* |
| *Achromobacter aloeverae* | *Aquitalea pelogenes* | *Balnearium lithotrophicum* | *Alicyclobacillus sendaiensis* | *Crenotalea thermophila* | *Fourniella massiliensis* | *Laceyella sediminis* | *Thermus brockianus* |
| *Acidimicrobium ferrooxidans* | *Arcobacter canalis* | *Belliline a caldifistulae* | *Alicyclobacillus shizuokensis* | *Deferribacter autotrophicus* | *Garciella nitratireducens* | *Lebetimonas natsushimae* | *Thermus caldifontis* |
| *Acidithiobacillus caldus* | *Arcobacter cibarius* | *Bogoriella caseilytica* | *Alicyclobacillus tengchongensis* | *Deferribacter desulfuricans* | *Geobacillus galactosidasius* | *Levilinea saccharolytica* | *Thermus caliditerrae* |
| *Acidobacterium ailaaui* | *Arcobacter haliotis* | *Brevibacillus borstelensis* | *Alicyclobacillus vulcanalis* | *Deferrisoma camini* | *Geobacillus icigianus* | *Lihuaxuella thermophila* | *Thermus filiformis* |
| *Acidothermus cellulolyticus* | *Ardenticatena maritima* | *Brockia lithotrophica* | *Alloiococcus otitis* | *Defluviitalea raffinosedens* | *Geobacillus jurassicus* | *Limisphaera ngatamarikiensis* | *Thermus igniterrae* |
| *Acinetobacter celticus* | *Athalassotoga saccharophila* | *Caenibacillus caldisaponilyticus* | *Alteribacillus persepolensis* | *Desulfacinum hydrothermale* | *Geobacillus kaustophilus* | *Litorilinea aerophila* | *Thermus oshimai* |
| *Acinetobacter colistiniresistens* | *Aurantimonas coralicida* | *Caldalkalibacillus thermarum* | *Alteromonas addita* | *Desulfacinum infernum* | *Geobacillus stearothermophilus* | *Lysinibacillus sphaericus* | *Thermus parvatiensis* |
| *Acinetobacter defluvii* | *Aureimonas altamirensis* | *Caldanaerobacter subterraneus* | *Alteromonas aestuarivivens* | *Desulfosoma caldarium* | *Geobacillus subterraneus* | *Mahella australiensis* | *Thermus scotoductus* |
| *Acinetobacter equi* | *Bacillus alcalophilus* | *Caldanaerobius fijiensis* | *Alteromonas genovensis* | *Desulfothermus okinawensis* | *Geobacillus thermanticus* | *Marinithermus hydrothermalis* | |
| *Acinetobacter seifertii* | *Bacillus altitudinis* | *Caldanaerobius polysaccharolyticus* | *Alteromonas oceani* | *Desulfotomaculum australicum* | *Geobacillus thermocatenulatus* | *Meiothermus chliarophilus* | |
| *Actinobaculum suis* | *Bacillus alveayuensis* | *Caldanaerovirga acetigignens* | *Alteromonas stellipolaris* | *Desulfotomaculum carboxydivorans* | *Geobacillus thermodenitrificans* | *Moorella glycerini* | |
| *Actinomadura formosensis* | *Bacillus amyloliquefaciens* | *Calderihabitans maritimus* | *Aminiphilus circumscriptus* | *Desulfotomaculum hydrothermale* | *Geobacillus thermoleovorans* | *Moorella humiferrea* | |
| *Actinomadura rubrobrunea* | *Bacillus aquimaris* | *Caldibacillus debilis* | *Ammonifex degensii* | *Desulfotomaculum nigrificans* | *Geobacillus toebii* | *Moorella mulderi* | |
| *Actinomyces ruminicola* | *Bacillus atrophaeus* | *Caldicellulosiruptor acetigenus* | *Ammonifex thiophilus* | *Desulfotomaculum profundi* | *Geobacillus vulcani* | *Moorella stamsii* | |
| *Actinopolyspora mortivallis* | *Bacillus azotoformans* | *Caldicellulosiruptor bescii* | *Amphibacillus jilinensis* | *Desulfotomaculum putei* | *Geobacillus zalihae* | *Moorella thermoacetica* | |
| *Aeribacillus composti* | *Bacillus bataviensis* | *Caldicellulosiruptor changbaiensis* | *Amphibacillus sediminis* | *Desulfotomaculum thermobenzoicum* | *Georgenia muralis* | *Natranaerobius thermophilus* | |
| *Aeribacillus pallidu* | *Bacillus cavernae* | *Caldicellulosiruptor hydrothermalis* | *Amphibacillus xylanus* | *Desulfotomaculum thermocisternum* | *Geosporobacter subterraneus* | *Natranaerobius trueperi* | |
| *Aeromonas bivalvium* | *Bacillus coagulans* | *Caldicellulosiruptor kristjanssonii* | *Amphiplicatus metriothermophilus* | *Desulfotomaculum thermosapovorans* | *Geothermobacter ehrlichii* | *Persephonella marina* | |
| *Aeromonas media* | *Bacillus cohnii* | *Caldicellulosiruptor kronotskiensis* | *Amycolatopsis eurythema* | *Desulfotomaculum thermosubterraneum* | *Halobacillus alkaliphilus* | *Phorcysia thermohydrogeniphila* | |
| *Aeromonas molluscorum* | *Bacillus cytotoxicus* | *Caldicellulosiruptor lactoaceticus* | *Amycolatopsis thermoflava* | *Desulfovirgula thermocuniculi* | *Heliobacterium modesticaldum* | *Pseudothermotoga thermarum* | |
| *Aeromonas sanarellii* | *Bacillus drentensis* | *Caldicellulosiruptor owensensis* | *Anaerobaculum hydrogeniformans* | *Desulfurella acetivorans* | *Herbinix hemicellulosilytica* | *Rhodothermus profundi* | |
| *Aeromonas simiae* | *Bacillus firmus* | *Caldicellulosiruptor saccharolyticus* | *Anaerobaculum mobile* | *Desulfurella amilsii* | *Herbivorax saccincola* | *Thermaerobacter marianensis* | |
| *Aeromonas taiwanensis* | *Bacillus fumarioli* | *Caldicoprobacter oshimai* | *Anaerobaculum thermoterrenum* | *Desulfurobacterium atlanticum* | *Hippea alviniae* | *Thermocrinis albus* | |
| *Afipia birgiae* | *Bacillus galactosidilyticus* | *Caldilinea aerophila* | *Anaerobranca californiensis* | *Desulfurobacterium indicum* | *Hippea jasoniae* | *Thermocrinis minervae* | |
| *Afipia massiliensis* | *Bacillus halosaccharovorans* | *Caldisalinibacter kiritimatiensis* | *Anaerobranca gottschalkii* | *Dictyoglomus thermophilum* | *Hippea maritima* | *Thermocrinis ruber* | |
| *Albidovulum inexpectatum* | *Bacillus indicus* | *Calditerricola satsumensis* | *Anaerolinea thermolimosa* | *Dictyoglomus turgidum* | *Hydrogenibacillus schlegelii* | *Thermodesulfatator autotrophicus* | |
| *Albidovulum xiamenense* | *Bacillus infantis* | *Calditerrivibrio nitroreducens* | *Anaerolinea thermophila* | *Dissulfuribacter thermophilus* | *Hydrogenivirga caldilitoris* | *Thermodesulfatator indicus* | |
| *Alcanivorax dieselolei* | *Bacillus jeotgali* | *Caldithrix abyssi Miroshni* | *Anaerophaga thermoh* | *Effusibacillus pohliae* | *Hydrogenobacter thermop* | *Thermosipho atlanticus* | |

| | | | chenko | alophila | | hilus | |
|---|---|---|---|---|---|---|---|
| *Alcanivorax nanhaiticus* | *Bacillus licheniformis* | *Caloramator fervidus* | *Anaerosalibacter bizertensis* | *Enterobacter mori* | *hydrogenophaga pseudoflava* | *Thermosulfurimonas dismutans* | |
| *Alicyclobacillus acidiphilus* | *Bacillus marisflavi* | *Caloramator mitchellensis* | *Ancylobacter pratisalsi* | *Enterococcus caccae* | *Hydrogenophaga taeniospiralis* | *Thermotoga maritima* | |
| *Alicyclobacillus acidocaldarius* | *Bacillus methanolicus* | *Caloramator proteoclasticus* | *Aneurinibacillus danicus* | *Enterococcus crotali* | *Hydrogenothermus marinus* | *Thermotoga naphthophila* | |
| *Alicyclobacillus acidoterrestris* | *Bacillus mojavensis* | *Caloranaerobacter azorensis* | *Aneurinibacillus thermoaerophilus* | *Fervidicella metallireducens* | *Inmirania thermothiophila* | *Thermotoga neapolitana* | |
| *Alicyclobacillus contaminans* | *Bacillus novalis* | *Caloranaerobacter ferrireducens* | *Anoxybacillus amylolyticus* | *Fervidicola ferrireducens* | *Keratinibaculum paraultunense* | *Thermovenabulum gondwanense* | |
| *Alicyclobacillus herbarius* | *Bacillus smithii* | *Caminibacter mediatlanticus* | *Anoxybacillus flavithermus* | *Fervidobacterium changbaicum* | *Kosmotoga arenicorallina* | *Thermovibrio ammonificans* | |
| *Alicyclobacillus hesperidum* | *Bacillus sonorensis* | *Caminicella sporogenes* | *Anoxybacillus pushchinoensis* | *Fervidobacterium gondwanense* | *Kosmotoga olearia* | *Thermovibrio guaymasensis* | |
| *Alicyclobacillus kakegawensis* | *Bacillus tequilensis* | *Clostridium clariflavum* | *Anoxybacillus tepidamans* | *Fervidobacterium islandicum* | *Kosmotoga pacifica* | *Thermovirga lienii* | |
| *Alicyclobacillus macrosporangiidus* | *Bacillus thermoamylovorans* | *Clostridium straminisolvens* | *Anoxybacillus thermarum* | *Fervidobacterium nodosum* | *Kroppenstedtia pulmonis* | *Thermus amylolyquefaciens* | |
| *Alicyclobacillus montanus* | *Bacillus thermocopriae* | *Clostridium thermocellum* | *Anoxybacillus vitaminiphilus* | *Fervidobacterium pennivorans* | *Kyrpidia spormanii* | *Thermus antranikianii* | |